\documentclass[twocolumn,secnumarabic,amssymb, nobibnotes, aps, prd]{revtex4}
\usepackage{graphicx}

\begin{document}
\title{Segment of an inhomogeneous mesoscopic loop as a dc power source}
\author{S.V. Dubonos, V.I. Kuznetsov, and A.V. Nikulov}
\affiliation{Institute of Microelectronics Technology and High Purity Materials, Russian Academy of Sciences, 142432 Chernogolovka, Moscow District, RUSSIA.} 
\begin{abstract} A dc voltage changed periodically with magnetic field is observed on segments of asymmetric aluminum loop without any external dc current at temperatures corresponded to superconducting transition. According to this experimental result a segment of the loop is a dc power source. A possibility of a persistent voltage on segments of an
inhomogeneous normal metal mesoscopic loop follows from this result. 
 \end{abstract}

\maketitle

\narrowtext

\section*{Introduction}

It is known that the persistent current (i.e. a direct current in
the thermodynamic equilibrium state) can flow along a mesoscopic loop
because of the quantization of the momentum circulation
$\int_{l} dl p = \int_{l} dl  (mv + (q/c)A) = m \int_{l} dl v + (q/c)\Phi = n2\pi \hbar $
The persistent current $I_{p} = s j_{p}$ in normal metal mesoscopic loops
was predicted more than 30 years ago [1] and was observed not so long ago [2].
It is known also that a potential difference $V = (<\rho>_{l_{s}} -
<\rho>_{l})l_{s}j$ should be observed on a segment $l_{s}$ of an
inhomogeneous conventional loop at a current density $j$ along the loop if the average
resistivity along the segment $<\rho>_{l_{s}} = \int_{l_{s}} dl
\rho/l_{s}$ differs from the one along the loop $<\rho>_{l} = \int_{l} dl \rho/l$.
Therefore a possibility of a persistent voltage can be assumed at a segment
of an inhomogeneous mesoscopic loop at $j_{p} \neq 0$.

The latter, i.e. $j_{p} \neq 0$, can be only if
the mean free path of electrons is not smaller than
the length $l$ of loop circumference and the temperature
is lower than the energy difference between adjacent
permitted states $p^{2}(n=1)/2m - p^{2}(n=0)/2m
= 2\pi^{2} \hbar^{2}/ml^{2} $. This difference is not large for
electrons. For example, at $l \approx 4 \ \mu m$,
the $2\pi^{2} \hbar^{2}/ml^{2} $ value corresponds
$T \approx 1 \ K$. Therefore it is enough difficult to observe
the persistent current in normal metal loop [2].

It is more easier to observe the persistent current in
superconducting loop since the mean
free path of superconducting pairs is infinite and
the energy difference between adjacent permitted states is
much higher than in normal metal loop.
First experimental evidence of  the persistent current at
non-zero resistance - the Little-Parks experiment [3] was made
as long ago as 1962.
In the present work the dc voltage proportional to the $I_{p}$
is observed on segments of an asymmetric superconducting
loop in accordance with the analogy with a conventional
inhomogeneous loop.

\begin{figure}[]
\includegraphics{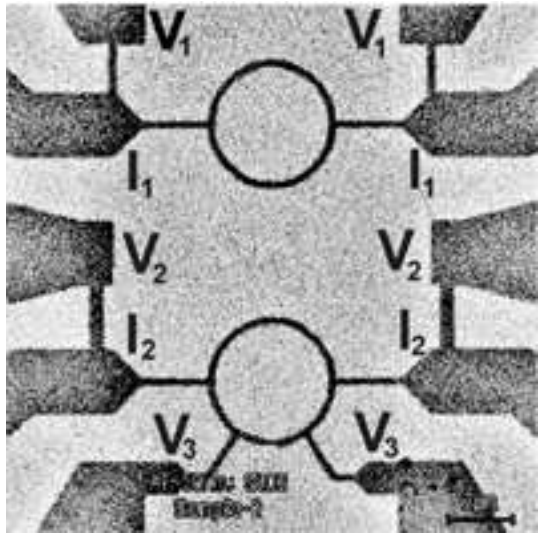}
\caption{\label{fig:epsart} An electron micrograph one of the aluminum loop
samples.  $I_{1}$ and $V_{1}$ are the current and potential contacts of
the  symmetric loop. $I_{2}$ and $V_{2}$ are the current and potential
contacts of the asymmetric loop. $V_{3}$ are the additional potential
contacts of the asymmetric loop.}
\end{figure}

\section{Experimental details}
The dependencies of the dc voltage $V$ on the magnetic
flux $\Phi \approx BS$ of some round symmetric and asymmetric Al loops
(see Fig.1) with a diameter $2r = l/\pi $ = 1, 2 and 4
$\mu m$ and a linewidth w = 0.2 and 0.4 $\mu m$ at the dc measuring current
$I_{m}$ and different temperature close to $T_{c}$ were measured. Here $B$ is the
magnetic induction produced by a superconducting coil; $S = \pi r^{2}$ is the area of the
loop. The Al microstructures Fig.1 are prepared using an
electron lithograph developed on the basis of a
JEOL-840A electron scanning microscop. The sheet resistance $R_{\diamond } \approx 0.5 \
\Omega /\diamond $ at 4.2 K, the resistance ratio $R(300 K)/R(4.2 K)
\approx 2$ and the midpoint of the superconducting resistive transition
$T_{c} \approx  1.24 \ K$.

\section{Experimental results}
The voltage oscillations corresponded to the conventional
Little-Parks oscillations were
observed on the contacts $V_{1}$ of the symmetric loop [4].
These resistance oscillations $R_{1}(\Phi/\Phi_{0}) = V_{1}/I_{1}$
observed in the temperature region of the superconducting transition
(i.e. at $T \approx T_{c}$) are explained by the $T_{c}$ oscillations
$T_{c}(\Phi/\Phi_{0})$ because of the oscillation of the persistent
current $I_{p}(\Phi/\Phi_{0})$ [5]. Here $\Phi_{0} = 2\pi \hbar c/q =
\pi \hbar c/e$ is the flux quantum for superconducting pair, $q = 2e$.
The voltage $V_{1} = 0$ at the measuring current
$I_{1} \equiv I_{m} = 0$ whereas the voltage oscillations
$V(\Phi/\Phi_{0})$ are observed on the contacts $V_{2}$ and $V_{3}$
of the asymmetric loop at $I_{2} \equiv I_{m} = 0$ Fig.2.

The apparent (with amplitude $\Delta V \geq 0.1 \mu V$) oscillations
without an external dc current were observed in a narrow  temperature
region $\Delta T \approx  0.01 \ K$ corresponds to the bottom of the resistive
transition. Its amplitude increases (up to $\Delta V \simeq 1.2 \mu V$)
with temperature lowering down to the lowest temperature
$\approx 1.23 \ K$ we could reach.

\begin{figure}[]
\includegraphics{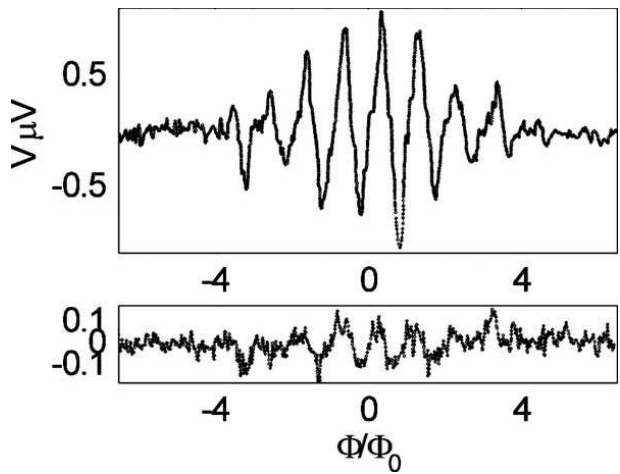}
\caption{\label{fig:epsart} Oscillation of the voltage measured on the $V_{2}$ contacts (upper curve) and on the $V_{3}$ contacts  (lower curve) of the asymmetric loop with 2r = 4 $\mu m$ and w = 0.4 $\mu m$. $I_{m} = 0$. $T = 1.231 K$ corresponded to the bottom of the resistive transition. }
\end{figure}

\section{Power source}
The observation of the voltage oscillations in the asymmetric
loop and its absence in the symmetric loop conform to the
analogy with a conventional loop. Although the oscillations
$I_{p}(\Phi/\Phi_{0})$ [5] take place in the both cases in the
symmetric loop $<\rho>_{l_{s}} = <\rho>_{l}$ and therefore
$V = 0$ at $I_{m} = 0$. The difference of $<\rho>_{l_{s}}$ from
$<\rho>_{l}$ in the asymmetric loop is caused by the
additional potential contacts $V_{3}$.

There is an important difference from the conventional
loop where the potential difference appears in accordance
with the Ohm's law $E = -\bigtriangledown V - (1/c)dA/dt = \rho j$.
The voltage on Fig.2 is observed without the Faraday's voltage
$dA/dt = (1/l)d\Phi/dt = 0$ and consequently the electric
field $E = -\bigtriangledown V$ and the persistent current $I_{p}$
should have opposite directions in a segment because $\int_{l} dl
\bigtriangledown V \equiv 0$, i.e. according to the result presented
on Fig.2 a segment of the asymmetric loop is a dc power source at
$\Phi \neq n\Phi_{0}$ and  $\Phi \neq (n+0.5)\Phi_{0}$. The power
$W_{load} =  V^{2}R_{load}/(R_{load}+R_{s})^{2}$ can be obtained
on a load with the resistance $R_{load}$. Because
the resistance of the segment $R_{s} \leq R_{sn} \approx 15 \
\Omega $ then  $W_{load} =
V^{2}/4R_{s} \geq  2 \ 10^{-14} \ Wt$ at $R_{load} = R_{s}$ and $V
 \approx 1 \ \mu V$.

\section{What energy is transformed in the power $VI_{p}$?}
It should be noted that already the classical Little-Parks experiment is
evidence of the dc power source. According to this experiment the
persistent current $I_{p} \neq 0$ is observed at non-zero resistance $R_{l} > 0$
along the loop and consequently an energy dissipation with the power
$R_{l}I_{p}^{2}$ takes place. The persistent current is maintained in spite
of this dissipation because of reiterated changes of the momentum
circulation of superconducting pairs at switching of the loop between
superconducting states with different connectivity [6].

Because of the
quantization the momentum circulation of pair $\int_{l} dl p =
2m \int_{l} dl v + (2e/c)\Phi $ changes from  $(2e/c)\Phi $ to $n2\pi \hbar $ at
each closing of superconducting state. The reiterated changes $n2\pi \hbar -
(2e/c)\Phi = 2\pi \hbar(n - \Phi/\Phi_{0})$ with an average frequency $\omega $ is
equivalent of the action of an average force $\int_{l} dl F_{q} = 2\pi \hbar
(\overline{n} - \Phi/\Phi_{0}) \omega $ which maintains the circulating current instead of
the Faraday's voltage. $\overline{n}$ is the thermodynamic average of the quantum
number $n$. $\overline{n} - \Phi/\Phi_{0} \approx (n - \Phi/\Phi_{0})_{min}$ when
$\Phi$ is not close to $(n+0.5)\Phi_{0}$ and $\overline{n} - \Phi/\Phi_{0} = 0$ at
$\Phi = (n+0.5)\Phi_{0}$, where the integer number $n$ in $(n - \Phi/\Phi_{0})_{min}$
corresponds to a minimum possible value $|n - \Phi/\Phi_{0}|$.

The quantum force $F_{q}$ is not localized in principle in any loop segment [6], i.e.
 $F_{q} = \int_{l} dl F_{q}/l $. Therefore a potential difference
$V = (\pi \hbar  \omega/e)(\overline{n} - \Phi/\Phi_{0}) (l_{s}/l) $
should be observed on a segment $l_{s}$ remaining in superconducting
state when other segment is switched in normal state with the frequency
$\omega $. This relation explains the observed voltage oscillation Fig.2.
Loop segments are switched in normal state at $T \simeq T_{c}$ by
thermal fluctuations and an external electric noise. Consequently
the energy of thermal fluctuations or an external electric noise
is transformed in the power $VI_{p}$ observed in our work.

\section{Conclusion}
The observation of the voltage oscillations Fig.2
is evidence of a possibility of an analogous observation on
segments of inhomogeneous normal metal mesoscopic loop.
There is an important question: thermal fluctuations or an external
noise induce the dc voltage. The temperature dependence
of the amplitude $\Delta V(T)$ observed in our work is evidence
of the latter. But it is enough obvious that thermal fluctuations can
also induce the dc voltage on segments of  inhomogeneous
mesoscopic loops [6].

\subsection*{Acknowledgements}
This work was financially supported by the Presidium of Russian Academy of Sciences in the Program "Low-Dimensional Quantum Structures"

\end{document}